\documentclass[10pt,a4paper]{article}
\pdfoutput=1 % if your are submitting a pdflatex (i.e. if you have images in pdf, png or jpg format)

\usepackage{lineno}
\usepackage{nameref}
\usepackage{graphicx}
\usepackage{cite}
\usepackage{authblk}
\usepackage{color}
\usepackage{amsmath}
\usepackage{geometry}
\usepackage{subcaption}

\begin{document}

\title{A measurement of the ionization efficiency of nuclear recoils in silicon}
\date{}

\author[ ]{
F.~Izraelevitch$^{2,4,}$\thanks{Corresponding author: fogo@fnal.gov},
D.~Amidei$^6$,
A.~Aprahamian$^7$,
R.~Arcos-Olalla$^5$,
G.~Cancelo$^4$,
C.~Casarella$^7$,
A.~E.~Chavarria$^3$,
P.~Collon$^7$,
J.~Estrada$^4$,
G.~Fern\'{a}ndez~Moroni$^4$,
Y.~Guardincerri$^4$,
G.~Guti\'{e}rrez$^4$,
A.~Gyurjinyan$^7$,
A.~Kavner$^6$,
B.~Kilminster$^8$,
A.~Lathrop$^4$,
J.~Liao$^8$,
Q.~Liu$^7$,
M.~L\'{o}pez$^1$,
J.~Molina$^1$,
P.~Privitera$^3$,
M.~A.~Reyes$^5$,
V.~Scarpine$^4$,
K.~Siegl$^7$,
M.~Smith$^7$,
S.~Strauss$^7$,
W.~Tan$^7$,
J.~Tiffenberg$^4$ 
and L.~Villanueva$^5$.
\small} 

\affil[1]{Facultad de Ingenier\'{i}a, Universidad Nacional de Asunci\'{o}n, Asunci\'{o}n, Paraguay}
\affil[2]{Departamento de F\'{i}sica, FCEN - Universidad de Buenos Aires, Buenos Aires, Argentina}
\affil[3]{Kavli Institute for Cosmological Physics and The Enrico Fermi Institute, The University of Chicago, Chicago, IL, United States of America}
\affil[4]{Fermi National Accelerator Laboratory, Batavia, IL, United States of America}
\affil[5]{Departamento de F\'{i}sica, Universidad de Guanajuato, Guanajuato, M\'{e}xico}
\affil[6]{Department of Physics, The University of Michigan, Ann Arbor, MI, United States of America}
\affil[7]{Department of Physics, University of Notre Dame, Notre Dame, IN, United States of America}
\affil[8]{Physik-Institut, Universit\"{a}t Z\"{u}rich, Z\"{u}rich, Switzerland}

\maketitle

\begin{abstract}
We have measured the ionization efficiency of silicon nuclear recoils with kinetic energy between 1.8 and 20~keV. We bombarded a silicon-drift diode with a neutron beam to perform an elastic-scattering experiment. A broad-energy neutron spectrum was used and the nuclear recoil energy was reconstructed using a measurement of the time of flight and scattering angle of the scattered neutron. The overall trend of the results of this work is well described by the theory of Lindhard \textit{et al.} above 4~keV of recoil energy. Below this energy, the presented data shows a deviation from the model. The data indicates a faster drop than the theory prediction at low energies. 
\end{abstract}

%%%%%%%%%%%%%%%%%%%
\section{Introduction}

The development of technologies for detecting low energy nuclear recoils has been a very active field in recent years, mainly driven by dark matter searches and coherent neutrino nucleus scattering (CENNS) experiments. When a nucleus recoils in a semiconductor detector, it loses its kinetic energy through two mechanisms: the generation of free charge carriers by ionization and the production of phonons by collisions with the lattice atoms. The partition of energy is quantified by the ionization efficiency, $\varepsilon$, defined as the ratio of the energy lost via ionization, $E_i$, to the kinetic energy of the nuclear recoil, $E_{NR}$. In the literature, $E_i$ is usually denoted by eV$_\text{ee}$ (for \textit{electron equivalent}, so-named because an electron recoil, because an electron recoil transforms all its kinetic energy in ionization), and $E_{NR}$ is quantified in eV$_\text{NR}$. A model for $\varepsilon$ in solids was developed in the 1960's by Lindhard \textit{et al.} \cite{cite:Lindhard}. A subsequent series of experiments found that this model correctly predicts $\varepsilon$ above 20~keV$_\text{NR}$ in several materials including silicon \cite{cite:Sattler}. In 1985, it was proposed that the hypothetical dark matter particles may interact with ordinary matter coherently generating nuclear recoils in the keV range \cite{cite:WIMPscat}. In order to calibrate the response of the semiconductor detectors used for these searches, a series of experiments measured $\varepsilon$ between 4 and 20~keV$_\text{NR}$ in silicon during the early 1990's. These measurements showed relatively good agreement with the Lindhard model \cite{cite:Gerbier, cite:Zecher, cite:Dougherty} in this energy range.

Strong constraints disfavoring the existence of high-mass dark matter by direct searches \cite{cite:LUX}, and the development of models that suggest the existence of low-mass dark matter particles have motivated the search for dark-matter particles with mass below 10 GeV \cite{cite:DMmodels1,cite:DMmodels2}. Novel detection techniques capable of measuring sub-keV nuclear recoils have made these searches possible. Current experiments using semiconductor detectors that measure nuclear recoils in the sub-keV range include COGENT \cite{cite:COGENT}, DAMIC \cite{cite:DAMIC}, EDELWEISS \cite{cite:EDELWEISS} and SuperCDMS \cite{cite:SuperCDMS}. These low-threshold experiments demand a new effort in nuclear recoil calibration at lower energies. In addition, experiments trying to detect the CENNS process, predicted by the standard model but never measured, also rely on the detection of low-energy nuclear recoils [13,14] \cite{cite:COHERENT, cite:CONNIE}. 

Ionization production by nuclear recoils in silicon has recently been measured in the energy range [0.7,~2]~keV$_\text{NR}$ using a photoneutron source \cite{cite:UCresults}. This result indicates for the first time a significant deviation from the Lindhard model. The energy range covered in Ref. \cite{cite:UCresults} does not overlap with the previous measurements that showed good agreement with Lindhard theory. In this work we present a measurement of the ionization efficiency of nuclear recoils in silicon performed with a neutron elastic-scattering experiment called Antonella. The presented result maps the transition between low-energy measurements \cite{cite:UCresults} and previous measurements, consistent with the Lindhard model at higher energies \cite{cite:Gerbier, cite:Zecher, cite:Dougherty}.

\begin{figure}[tb]
	\begin{center}
		\includegraphics[width=0.6\textwidth]{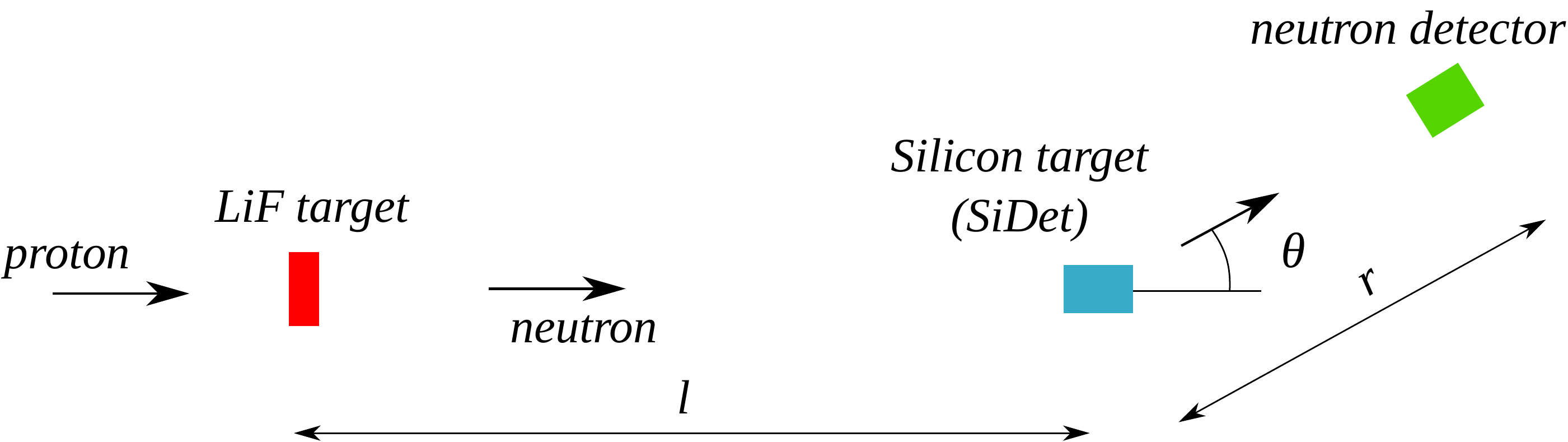}	\caption{Schematic layout of the experimental arrangement.}	\label{fig:ExpSchem}
	\end{center}
\end{figure}

%%%%%%%%%%%%%%%%%%%
\section{Experimental method}

%%%%%%%%%%%%%%%%%%%
\subsection{General description}

Figure~\ref{fig:ExpSchem} shows the schematic description of the experiment. Neutrons, produced by a proton beam impinging on a LiF target, scatter elastically from silicon nuclei in the sensitive bulk of a silicon detector (SiDet). The nuclear recoils deposit their kinetic energy in the SiDet producing an ionization signal ($E_i$), while the neutrons, scattered off the target, are detected again by a secondary neutron detector. The energy of the nuclear recoil in the SiDet can be calculated with non-relativistic kinematics as

\begin{equation} \label{eqKin1}
E_{NR} = E_n \frac{2}{(A+1)^2} \left[ A + \sin^2 \theta - \cos \theta \sqrt{ A^2-\sin^2\theta } \right] \, ,
\end{equation}

\noindent where $E_n$ is the energy of the incoming neutron, $\theta$ is the scattering angle with respect to the beam direction, and $A$ is the mass number of the silicon nucleus. Traditionally, these kind of experiments are done with monochromatic beams of known energy. However, it is possible to use a broad neutron energy spectrum and determine the neutron energy on an event-by-event basis using measurements of the time of flight (ToF) of the neutron and its scattering angle. The total ToF of the neutron from the neutron-production target to the neutron detector, $\Delta t$, is 

\[ \Delta t = \frac{l}{\sqrt{\frac{2E_n}{m}}} + \frac{r}{\sqrt{\frac{2E_s}{m} } }  \]

\noindent where $l$ is the distance from the neutron production target to the SiDet, $r$ is the distance from the SiDet to the neutron detector, $m$ is the mass of the neutron, and $E_s$ is the energy of the scattered neutron. The latter can be related to $E_n$ by

\[ E_s = E_n \frac{1}{(A+1)^2} \left( \cos\theta + \sqrt{A^2+\sin^2\theta} \right) ^2  \]

\noindent which yields an expression to calculate the energy of the incoming neutron from the geometry of the setup and the timing difference from the arrival at the target to the secondary neutron detector by:

\begin{equation} \label{eqKin2}
E_n = \frac{m}{2(\Delta t)^2} \left[ l + r \; \frac{A+1}{\cos\theta+\sqrt{A^2-\sin^2\theta}} \right]^2 \,.
\end{equation}

\begin{figure}[tb]
	\begin{center}
		\includegraphics[width=0.6\textwidth]{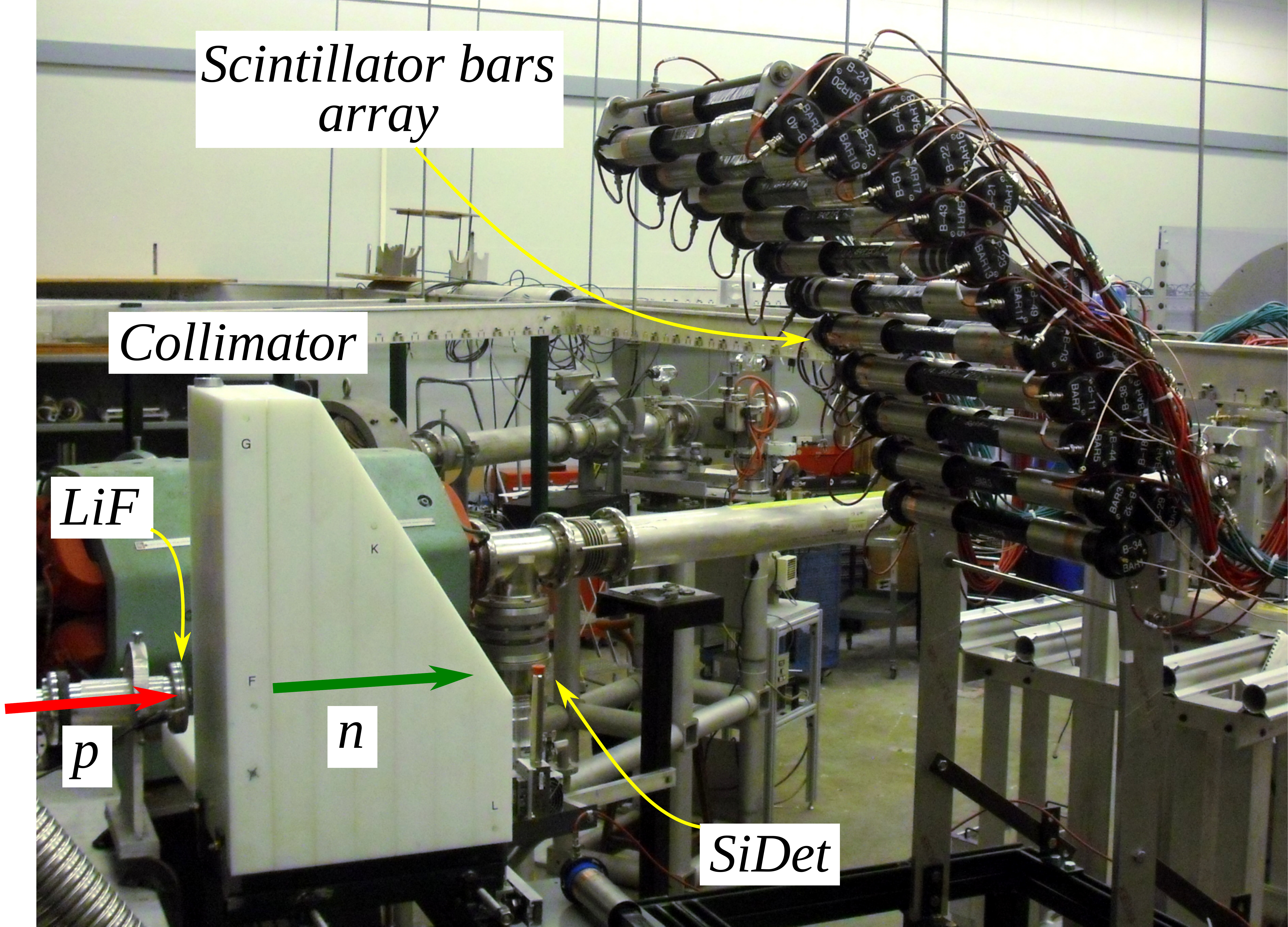}	\caption{Picture of the setup at the FN Van de Graaff at the ISNAP, UND.}	\label{fig:setupPic}
	\end{center}
\end{figure}

\noindent Thus, on an event-by-event basis, the nuclear recoil energy in the SiDet ($E_{NR}$) can be calculated from the ToF measurement and the scattering angle ($\theta$), with the use of Eqs. (\ref{eqKin1}) and (\ref{eqKin2}). The full experimental setup is shown in figure~\ref{fig:setupPic} and, in the next sections, individual components of the experiment are described. 

%%%%%%%%%%%%%%%%%%%
\subsection{Silicon detector}  \label{sec:SiDet}

\begin{figure}[t!]
	\begin{center}
		\includegraphics[width=0.7\textwidth]{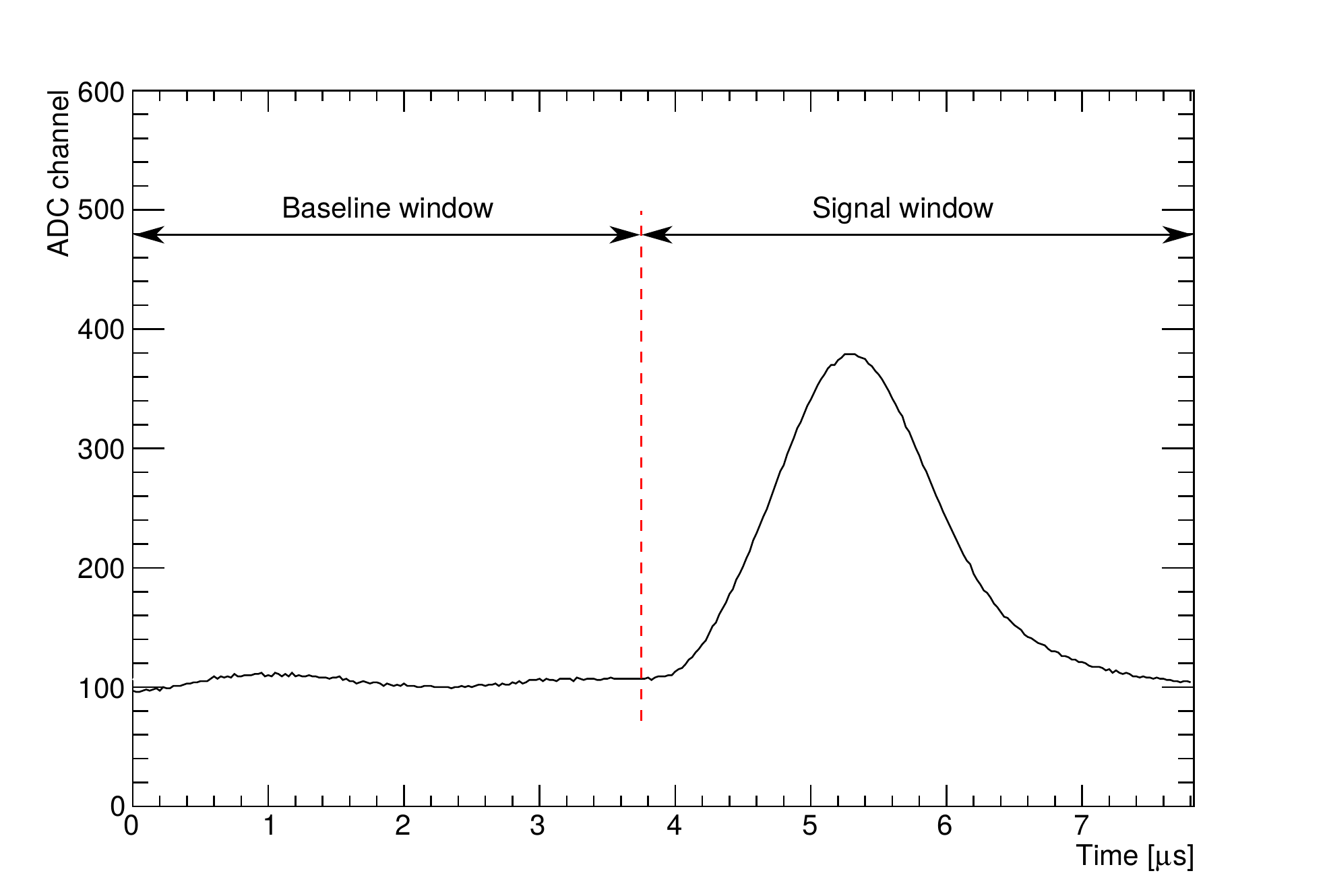}
		\caption{Example of a digitized SiDet waveform, acquired during the physics run. The figure shows the baseline and signal regions used to determined the ionization produced by the energy deposition. In this event, the pulse in the signal window indicates an energy deposition of 2.72~keV.}	\label{fig:SiDetWaveform}
	\end{center}
\end{figure}

A commercial X-ray detector was used (XR-100 SDD, Amptek), which consisted of a peltier-cooled silicon-drift diode with a reset-type preamplifier in the same housing. In a silicon-drift diode the charge released by ionizing particles moves towards a small-central anode by the action of an electric field generated by a set of concentric electrodes. The timing resolution is about one $\mu$s, given by the location of the hit in the SiDet and the corresponding drift time of the charge-carriers~\cite{cite:NASA_SDD}. The bias voltage of the detector and preamplifier were supplied by the vendor electronics (PX5, Amptek). The SiDet has a mass of $\approx 30$~mg and was operated at 220~K with a bias of 110~V. The output signal of the detector was shaped with a spectroscopy amplifier and acquired with a waveform digitizer. Figure~\ref{fig:SiDetWaveform} shows an example of a digitized SiDet waveform. To determine the energy deposited in ionization, the baseline and signal windows were integrated and subtracted (see section~\ref{sec:DAQ} for details on the data acquisition system and how the system was triggered). 

\begin{figure}[ht!]
	\begin{center}
		\includegraphics[width=0.7\textwidth]{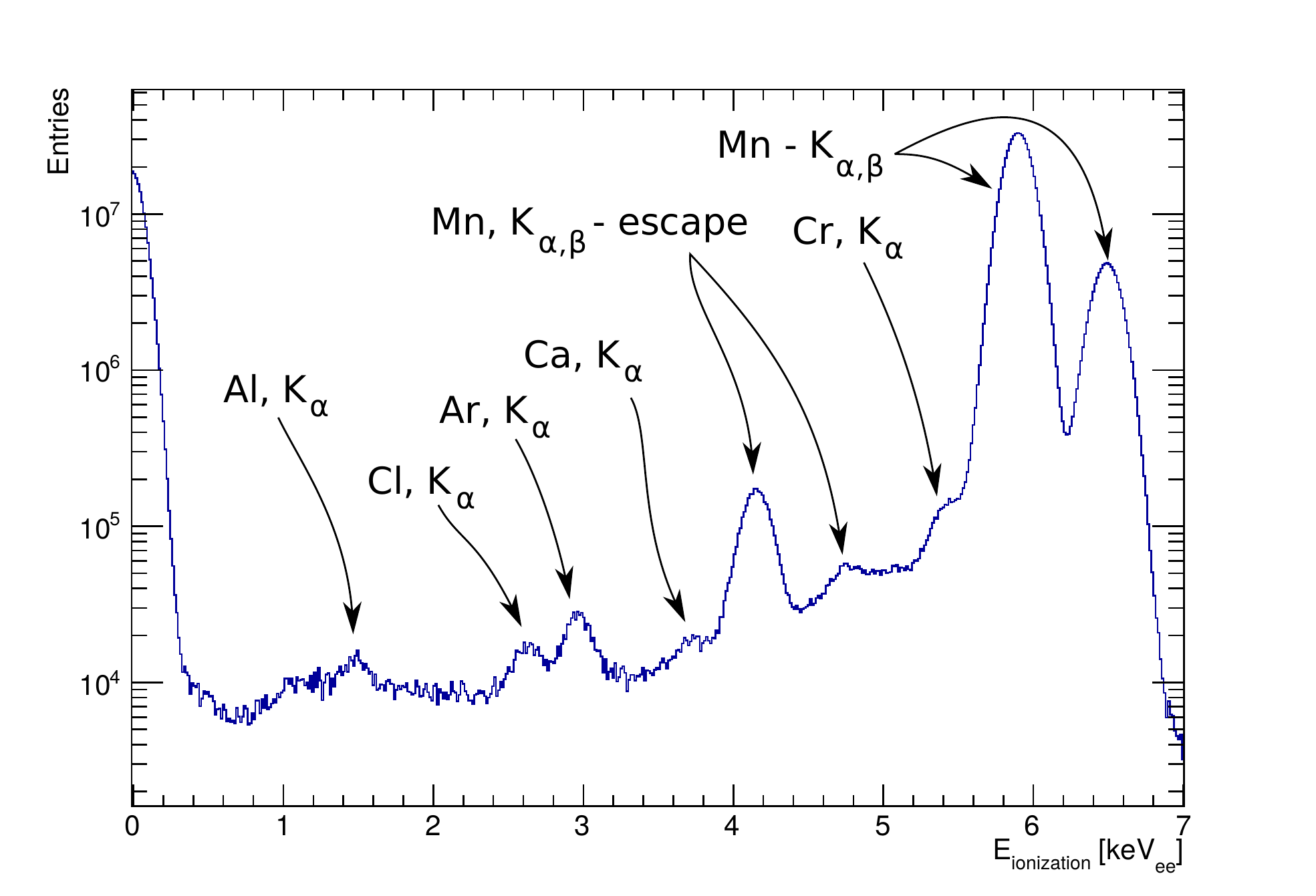}
		\caption{Energy spectrum recorded by the SiDet in a calibration with an $^{55}$Fe source. Several X-ray lines are identified (see text for details). The excess below 0.3~keV corresponds to the noise of the detector.}	\label{fig:SiDetCalibrated}
	\end{center}
\end{figure}

The SiDet was calibrated daily with an $^{55}$Fe source during the data taking, and the primary peak position of the spectrum was found to be stable to within 1~\%. Figure~\ref{fig:SiDetCalibrated} shows the energy spectrum obtained from one of these calibrations. The highest magnitude peaks correspond to X-ray lines of Mn K$_{\alpha}$ and K$_{\beta}$ produced in the $^{55}$Fe decay. The escape lines of Mn K$_{\alpha}$ and K$_{\beta}$ can also be seen. Also evident are the K$_{\alpha}$ lines from Cr, Ca, Ar, Cl and Al, from fluorescence in materials surrounding the silicon-drift diode. The Cr and Al materials are part of a multilayer collimator attached to the silicon chip (used in X-rays measurements applications). The Cl and Ca lines are present due to salt from fingerprints on the detector housing and Ar is in the air. 

\begin{figure}[t!]
	\begin{center}
		\includegraphics[width=0.7\textwidth]{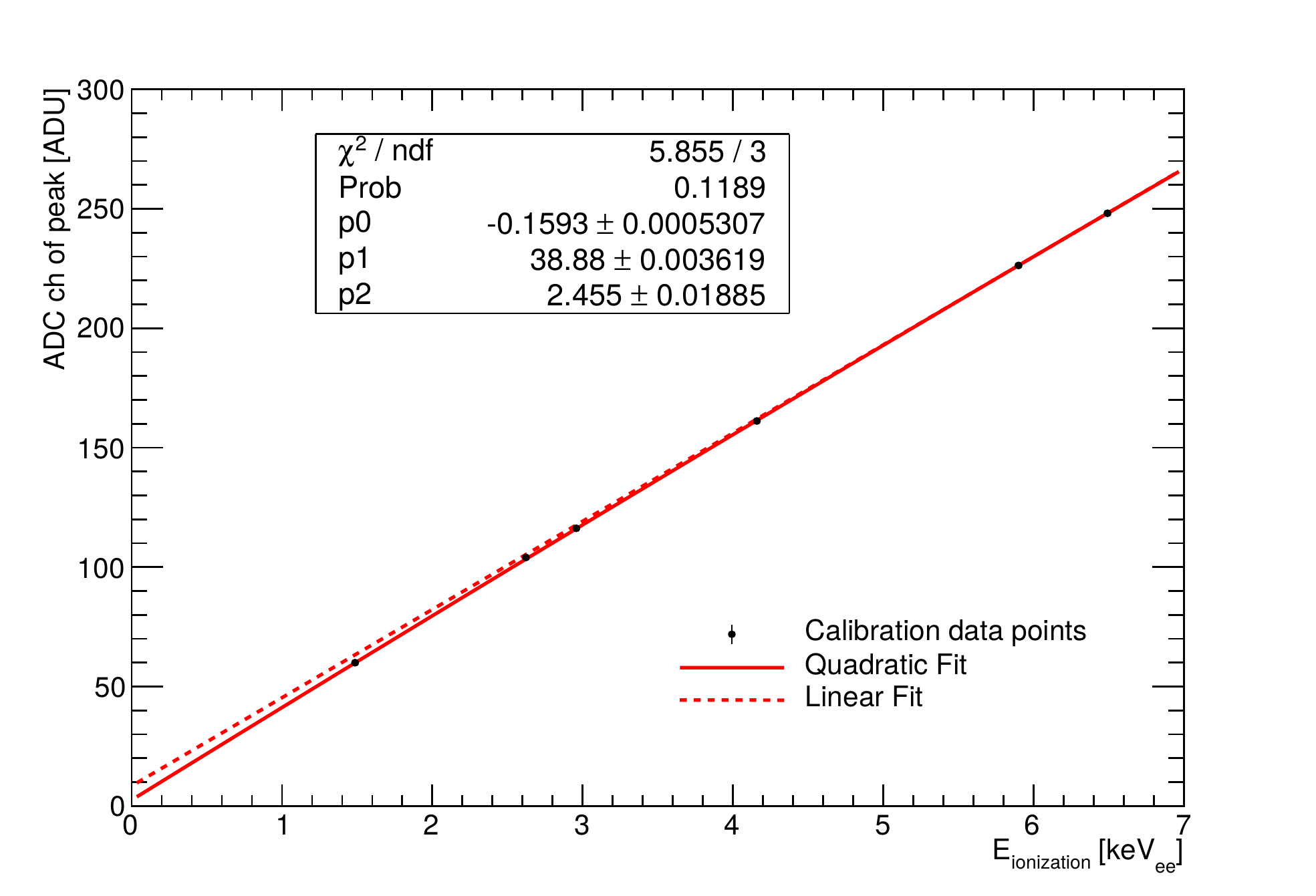}
			\label{fig:SiDetQuadCalibrationFullRange}
		\includegraphics[width=0.7\textwidth]{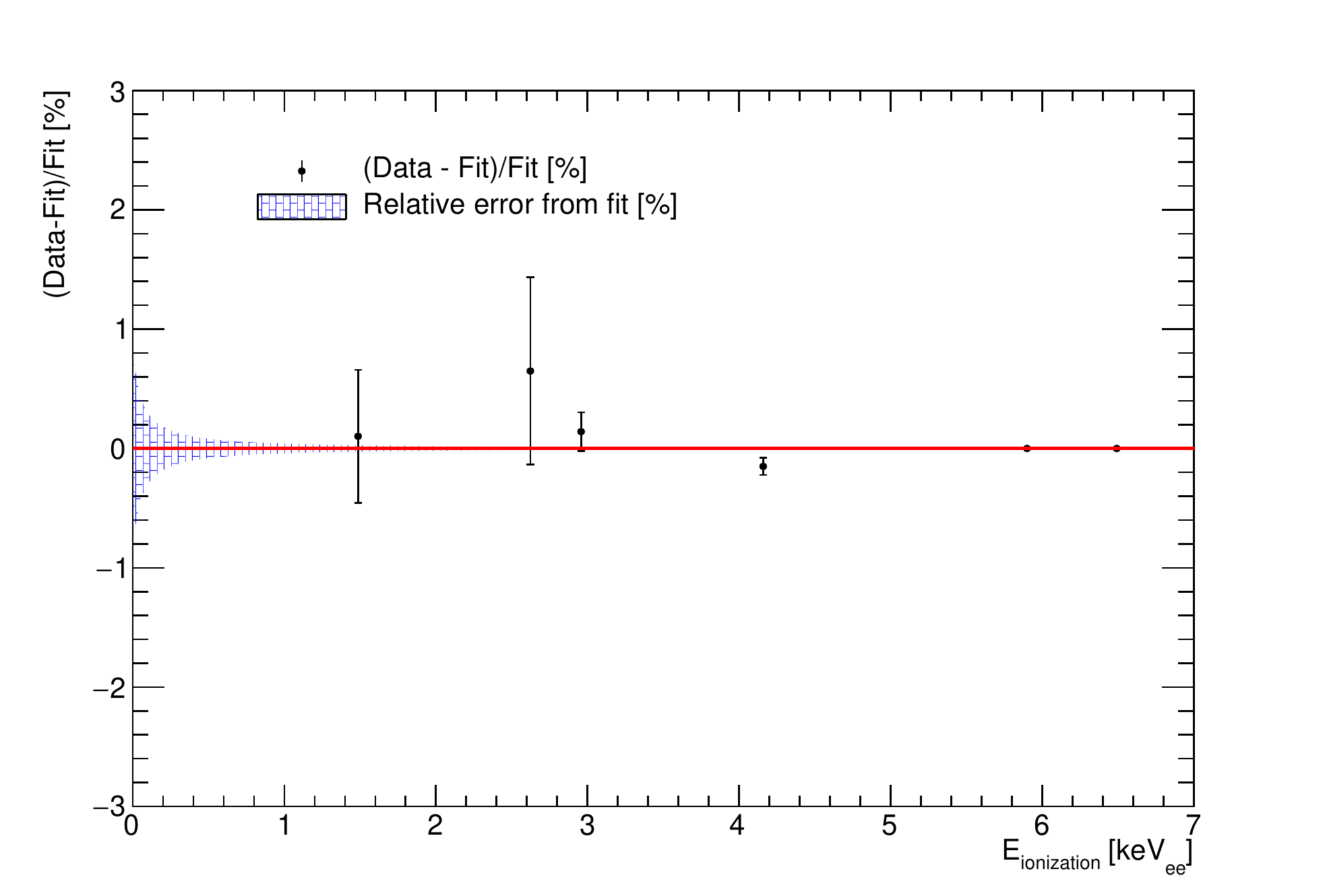}		
		\caption{Top panel: measured ADC channel for the six most prominent X-ray lines from the spectrum shown in figure~\ref{fig:SiDetCalibrated} as a function of the energy; the solid line is the quadratic fit performed used as a calibration; the dashed line is the linear fit used to evaluate the systematic uncertainty in the calibration; inset: fit results and goodness-of-the-fit estimators. Bottom panel: relative difference between the data and the fit (data points); relative error of the fit (shaded area).}	\label{fig:SiDetCalibration}
	\end{center}
\end{figure}

The most prominent peaks of the spectrum (Al-K$_\alpha$, Cl-K$_\alpha$, Ar-K$_\alpha$, Mn-K$_\alpha$, Mn-K$_\beta$ and escape line of Mn-K$_\alpha$) were used as calibration points of known energy and their centroids were determined individually by performing a fit of a Gaussian plus a linear function. The fit was done in a range of $\pm$ two-sigma around the Gaussian mean. The top panel of figure~\ref{fig:SiDetCalibration} shows the calibrations points in a scatter plot of the analog-to-digital converter (ADC) counts vs. energy.

In principle, the SiDet performance at low energies was demonstrated by the manufacturer down to 500~eV using X-rays, showing a linear relation between signal and energy~\cite{cite:AmptekIntNote}. In practice, a nonlinear behavior was observed at low energies that could be due to any of the components of the data acquisition system (digitizer and amplifier, described in section~\ref{sec:DAQ}). This non-linearity was quantified by adding a quadratic term to the calibration function ($\text{ADC ch} = p_0 E^2 + p_1 E + p_2$, $E$ in keV). The quadratic function was chosen because it minimized the $\chi^2$ among the simplest nonlinear models. The top panel of figure~\ref{fig:SiDetCalibration} shows the quadratic fit to the calibration points performed (solid line) with its statistical box, along with a linear fit (dashed line) for comparison. The bottom panel of figure~\ref{fig:SiDetCalibration} is the residual plot of the quadratic fit. It shows the relative difference between the data and the fit (data points), and the relative error obtained from the quadratic fit (shaded area). 

The silicon-drift diode was factory-sealed with a beryllium window of 13~$\mu$m. In order to repeat the manufacturer calibration below the energy of the Al-Kα with fluorescence lines, the irreversible removal of the beryllium window would have been required, making the detector unsuitable for the experiment. The systematic uncertainty due to the observed non-linearity at low energies was estimated as the difference between the quadratic and linear fits to the data, and was adopted in order to provide a conservative estimate of the low-energy calibration.

%%%%%%%%%%%%%%%%%%%
\subsection{Neutron production}
\label{sec:neutronProd}

\begin{figure}[t!]
	\begin{center}
		\includegraphics[width=0.7\textwidth]{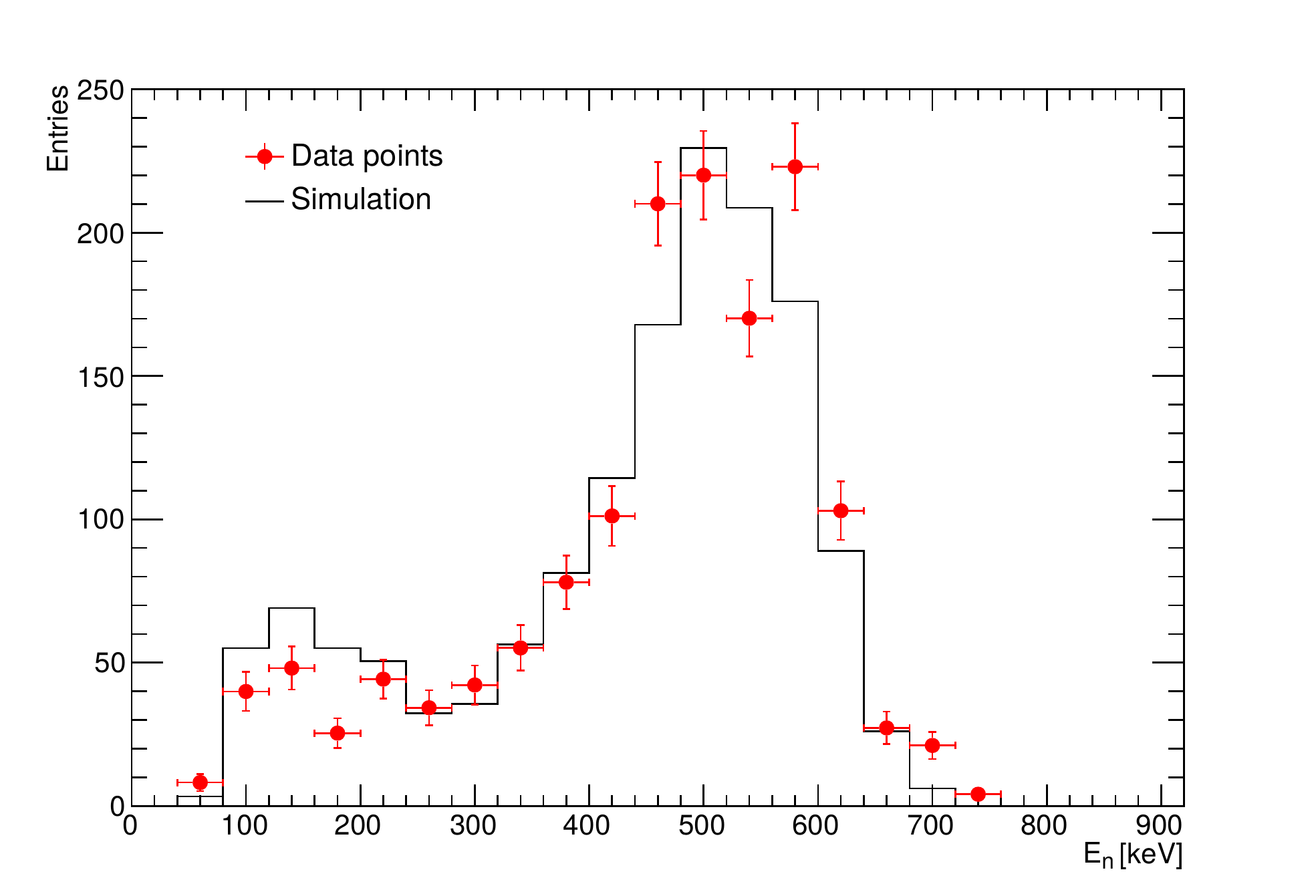}		
		\caption{Neutron spectrum produced with the $^7$Li(p,n)$^7$Be reaction using a 4.74~mg/cm$^2$ film of LiF. Solid circles: measurement using a scintillator bar in the beam axis and the ToF. Histogram: the simulated spectrum.}
		 	\label{fig:theNeutronSpectrum}
	\end{center}
\end{figure}

The experiment was held at the FN tandem Van de Graaff accelerator of the Institute for Structure and Nuclear Astrophysics (ISNAP), University of Notre Dame (UND), Indiana, U.S.A.. A $^7$Li(p,n)$^7$Be reaction at 0$^\circ$ was used \cite{cite:Burke}, with a proton beam energy of 2.326~MeV. The resolution of the accelerator buncher was $<$~2~ns. The accelerator bunch separation was set to 1~$\mu$s to be compatible with the drift time in the SiDet. The proton current was kept as high as possible, and averaged $\approx35$~nA. The target material was a 4.74~mg/cm$^2$ film of LiF, deposited on 197~mg/cm$^2$ of Au, onto an Al backing foil. The target was produced at Argonne National Laboratory, U.S.A.. The LiF thickness was optimized to maximize the number of neutrons produced as well as the number escaping the LiF. This yielded a broad-energy neutron spectrum in the range [0,600]~keV. In order to prevent the interaction of neutrons that travel directly from the LiF production target to the neutron detector without impinging on the SiDet, a high-density polyethylene collimator was interposed between the LiF target and the SiDet. The collimator hole was 5.5~mm of diameter and the maximum absorbing thickness was 35 cm. 

The neutron energy spectrum was characterized in a special run, where one scintillator bar (detector described in the next section) was placed in the neutron beam axis. The energy of the neutrons was determined event-by-event by measuring the ToF from the LiF production target to the scintillator bar. This run was also simulated in Geant4 (see section~\ref{sec:Simulation}). The expected neutron spectrum was calculated using the LiF target thickness and the energy dependence of the cross sections of the neutron production and detection ($^7$Li(p,n)$^7$Be and $^1$H(n,n')$^1$H, respectively \cite{cite:NNDC}). This calculated spectrum was used as an input in the simulation. Then the neutron transport to the detectors and their timing response was simulated. Finally, from the timing information obtained in the simulation, the neutron spectrum was reconstructed and compared with the measured one. Figure~\ref{fig:theNeutronSpectrum} shows the measured neutron spectrum and the one obtained in the simulation (the statistical error bars of the simulation are negligible). As shown, the measured neutron spectrum was well-reproduced by the simulation.

Neither the measured nor the simulated neutron spectrum was used in the measurement of the ionization efficiency in silicon, as the energy of each individual neutron was measured regardless of the spectrum. Nevertheless, this study verified that the salient features of the simulation were correct, allowing it to be used for evaluating systematic uncertainties and to understand and minimize the backgrounds (section~\ref{sec:Simulation}).

%%%%%%%%%%%%%%%%%%%
\subsection{Neutron detector}

The neutron detector was an array of 21 plastic scintillators bars (EJ-200, Eljen Technology) with a cross-section of 3$\times$3~cm$^2$  and length of 25~cm, coupled to two PMTs (EMI 9954KB, ET Enterprises), one at each end. The coupling was done with optical cement (EJ-500, Eljen Technology). The PMTs were refurbished from the CDF Central Hadronic Calorimeter \cite{cite:CDF}. Each PMT and base was tested and the malfunctioning ones were discarded. Several quantities were evaluated: voltage of all the electrodes of all of the bases, the dark current and dark pulse rate versus bias voltage, and the pulse amplitudes of 1 and $\approx 10$~p.e. obtained by illuminating the PMTs with a dim pulsed laser. The HV was individually adjusted to have the trigger at about 0.2~p.e. at $-30$~mV, the minimum of the edge discriminator level. A description of the scintillator bar assembly and a study of its response to low energy X-rays can be found in Ref. \cite{cite:Junhui}.

The neutron detector array covered a solid angle from 12.6$^\circ$ to 74.0$^\circ$ with respect to the beam axis. The bars were positioned in two layers, to pack them as closely as possible given the mechanical restrictions imposed by the existing PMT bases. The distance between the SiDet and the LiF neutron-production target was $l=51.1$~cm, with the collimator in between. The distance from the SiDet to the scintillator bars, $r$, was in the range between 80.0 and 88.9~cm, depending on the bar. The collimator, SiDet and neutron detector were mounted on a cart with the adjustments needed to position and align the system. The mechanical design was done in order to minimize the material on which neutrons could bounce and produce background. The geometry was surveyed by the Alignment and Metrology Department of Fermilab with a laser tracker (Radian, Automated Precision Inc.), with an instrumental accuracy of 10~$\mu$m \cite{cite:Survey}.

%%%%%%%%%%%%%%%%%%%
\subsection{Data acquisition system}
\label{sec:DAQ}

\begin{figure}[tb]
	\begin{center}
		\includegraphics[width=0.6\textwidth]{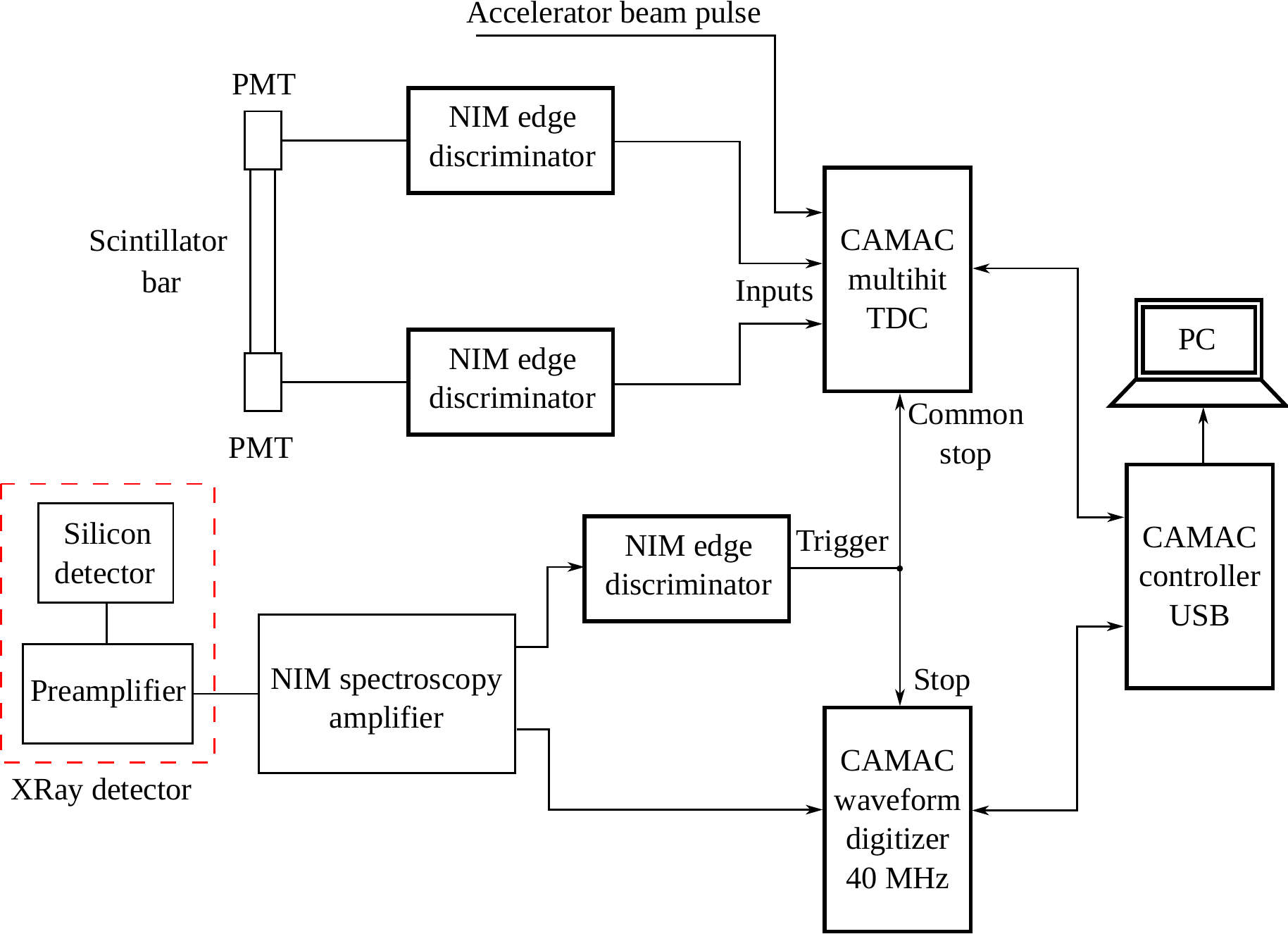}	\caption{Data acquisition block diagram.}	\label{fig:DAQschem}
	\end{center}
\end{figure}

The DAQ system was built with NIM and CAMAC modules. The block diagram is shown in figure~\ref{fig:DAQschem}. The anode signal of each PMT was connected to an edge discriminator (620AL, LeCroy), whose output was connected to a multi-hit TDC (3377, LeCroy). The TDC was used in Common Stop mode. When a Common Stop signal was applied to the module, it digitized the time of the hits that occurred within a time window, backward in time with respect to the Common Stop signal. The TDC range was set to 6.5~$\mu$s, to be able to track backgrounds from previous and subsequent bunches.

The output signal of the SiDet pre-amplifier was connected to a spectroscopy amplifier (2025, Canberra). One of the shaped outputs was connected to a CAMAC waveform digitizer of 40~MSPS (2262, LeCroy). This module worked in a pre-trigger mode. When a Stop signal was applied to the module, it digitized the last 313 samples in a window of 7.825~$\mu$s, backward in time with respect to the Stop signal. The first 150~samples were used to compute the baseline and the last 163 were used to compute the signal. Another shaped output of the spectroscopy amplifier was connected to an edge discriminator, which was responsible for producing the trigger signal of the whole DAQ system. The discriminator threshold level was set to trigger on the SiDet noise tail, at around 140~eV, to maximize the number of neutron events read out by the DAQ while keeping the dead time below 20~\%. The threshold level and trigger rate were monitored during the whole experiment. 

The trigger signal was delayed and connected to the Common Stop of the TDC and to the Stop signal of the waveform digitizer. The TDC and the digitizer were read out with a USB CAMAC controller (CC-USB, Wiener), which was connected to a computer for data storage. The accelerator beam pulse was connected to one of the TDC channels, and was the only signal between the experimental DAQ and the accelerator electronics. The zero time for each PMT channel, i.e. the time when the proton bunch hits the LiF target, was determined by measuring the arrival-time of the prompt gammas emitted in the $^7$Li(p,n)$^7$Be reaction \cite{cite:Fowler} along with the beam pulse. The experiment acquired data for 10 consecutive days, 24 hours a day except for planned interruptions to calibrate the SiDet with an $^{55}$Fe source. The trigger rate was about 170 Hz. The offline analysis showed that particles hit the SiDet at $\approx$ 4 Hz.

%%%%%%%%%%%%%%%%%%%
\subsection{Simulation}
\label{sec:Simulation}

The ionization efficiency experiment was simulated with Geant4 \cite{cite:GEANT4}. In the design stage, the simulations were used to assist the design of the experiment by identifying possible sources of background. As an example, the collimator was initially conceived as a rectangular prism. The simulations showed that there was a significant fraction of simulated neutrons that were back-scattered in the SiDet and then mirror-reflected in the downstream face of the collimator, back to the neutron detector. The collimator was then redesigned with a cut-out to reduced the solid angle subtended with respect to the SiDet, as shown in figure~\ref{fig:setupPic}. 

Simulations were also used in the analysis stage, running the analysis scripts both on the real data set and on the simulated one. The objective was to produce simulated data sets with an overall pattern as similar as possible to the real one. The geometry described in the simulation code was comprised of the collimator, the SiDet, the neutron detector including the scintillator bars and PMTs, and the structural hardware used in the mechanical support of the setup. The simulation generation started with the neutron spectrum (see section~\ref{sec:neutronProd}) and then computed the transport and kinematics of the neutrons. The simulation recorded the nuclear recoil kinetic energy and the timing information in the scintillator bars. Then, to approximate the function $E_i=f(E_{NR})$ that transforms the energy of the nuclear recoil to the energy of ionization (the simulated $\varepsilon$), a fit to the data was performed using a parametrization with a power law. Finally, on an event-by-event basis, we introduced fluctuations in the ionization process from nuclear recoils by adding Gaussian smearing with variance $\sigma^2(E_i) = 0.0125\, E_i$, following the calculation of Ref. \cite{cite:Lindhard}. 

The simulations were not used to obtain mean values of the data points of the ionization efficiency measurement, but to evaluate systematic uncertainties on them. The procedure was to modify one or more parameters in the simulation and to quantify the effect in the result obtained by running the analysis script on the simulated data set, comparing the modified simulation with the unmodified one.

%%%%%%%%%%%%%%%%%%%
\section{Results}

\begin{figure}[t!]
	\begin{center}
		\includegraphics[width=0.7\textwidth]{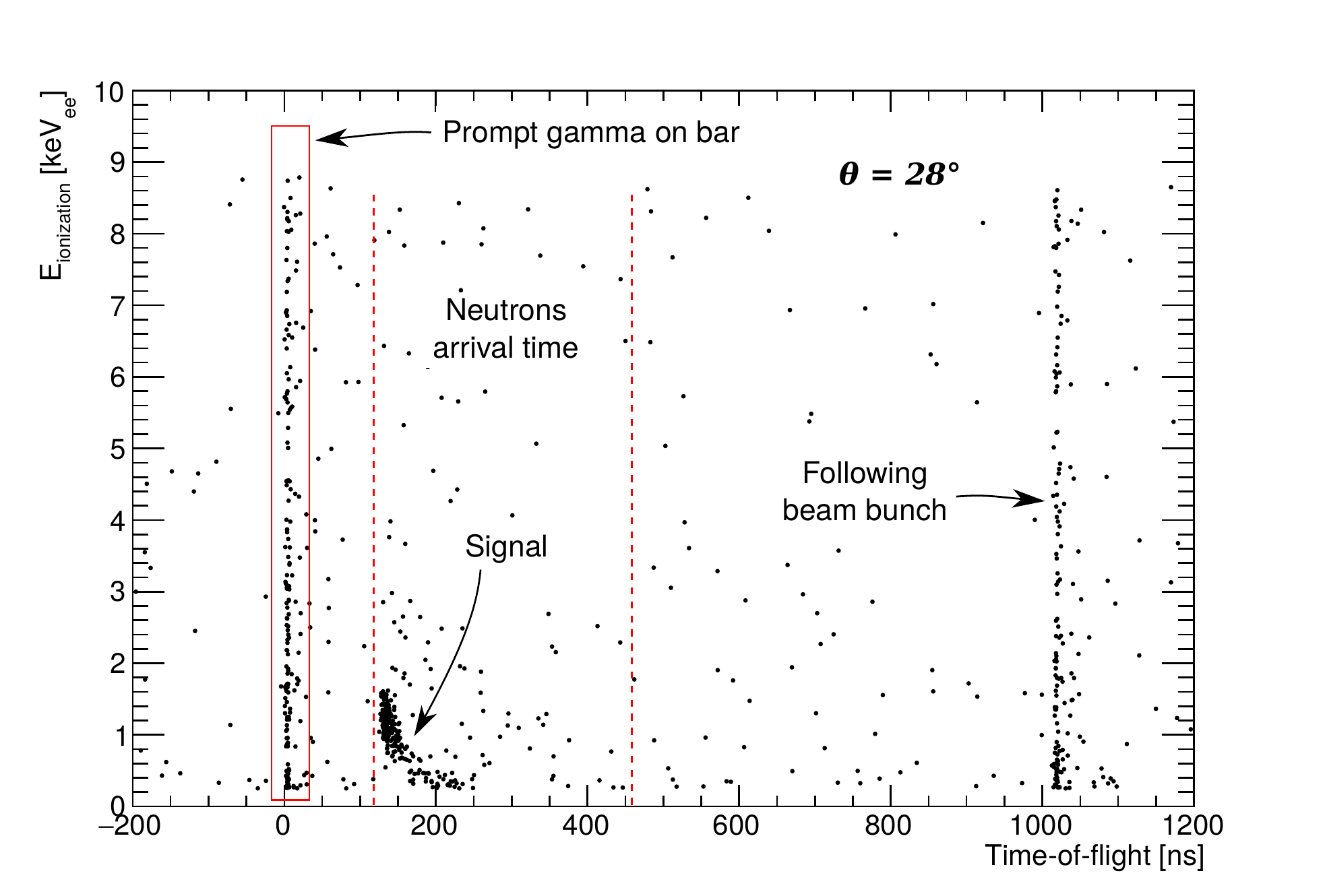}	
		\includegraphics[width=0.7\textwidth]{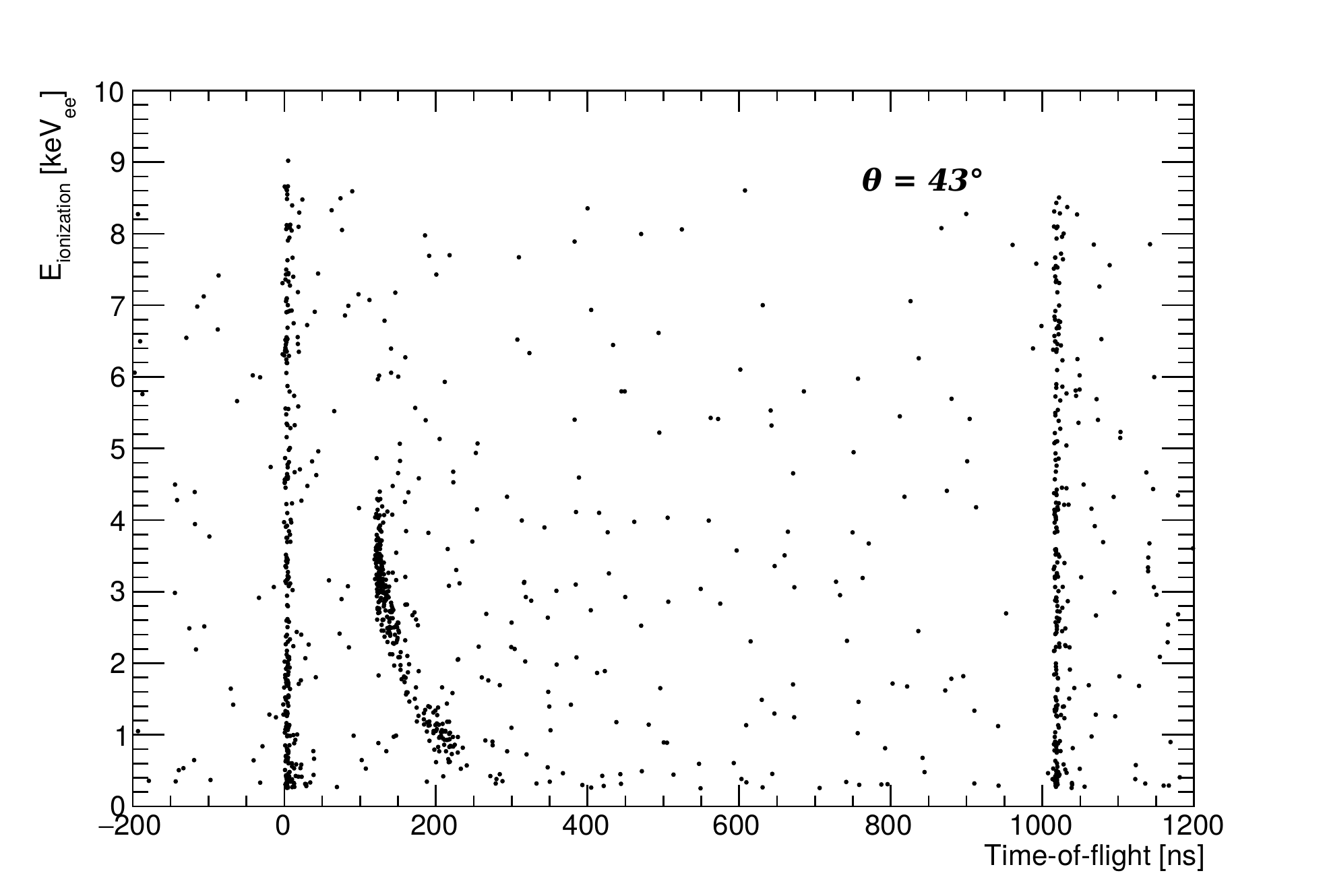}  \caption{Data showing ionization energy vs. ToF, for hits measured in the bar located at 28$^\circ$ from the beam line. The signal pattern and different backgrounds are identified, as described in the text. The bottom plot shows data collected in the bar at 43$^\circ$ for comparison.}	\label{fig:dataEionVsToF}
	\end{center}
\end{figure}

Figure~\ref{fig:dataEionVsToF} shows the correlation of the experimental parameters registered event-by-event: the ionization energy recorded in the SiDet, $E_i$, as a function of the total ToF determined for the neutron. The data points in the top and bottom plots are data for scattering angles of 28$^\circ$ and 43$^\circ$, respectively.

The total ToF is the time from the neutron production, i.e. when the protons hit the LiF target, until the neutron hits a scintillator bar. In the graphs, the vertical accumulation of points around 4~ns is due to accidental coincidences between prompt gammas hitting a bar and another particle hitting the SiDet, filling the whole energy range. This is consistent with the approximate 1.2~m distance between the LiF foil and the scintillator bars. The neutron arrival time is given by the geometry and the kinematics. It is, depending on the bar considered, between 120 and 450~ns for neutrons between 50 and 600~keV. The gammas produced in the following beam bunch can be seen at $\approx 1019$~ns, consistent with the accelerator bunch frequency of 985.5~kHz. These events include all the particles hitting the SiDet and prompt gammas of the following bunch hitting a scintillator bar. 

The events of interest for the nuclear recoil efficiency measurement are in the crescent-moon pattern labelled as \textit{signal} in the top plot of figure~\ref{fig:dataEionVsToF}. Events in the signal region behave approximately as $1/t^2$, as expected from the kinematic equations (\ref{eqKin1}) and (\ref{eqKin2}).  In both plots, the signal region shows a depleted zone in the region from approximately 150 to 170 ns which is consistent with the convolution of the 600-keV resonance of the neutron production \cite{cite:Burke} and the 200-keV resonance of the silicon elastic-scattering cross-sections \cite{cite:NNDC}. 

\begin{figure}[t!]
	\begin{center}
		\includegraphics[width=0.7\textwidth]{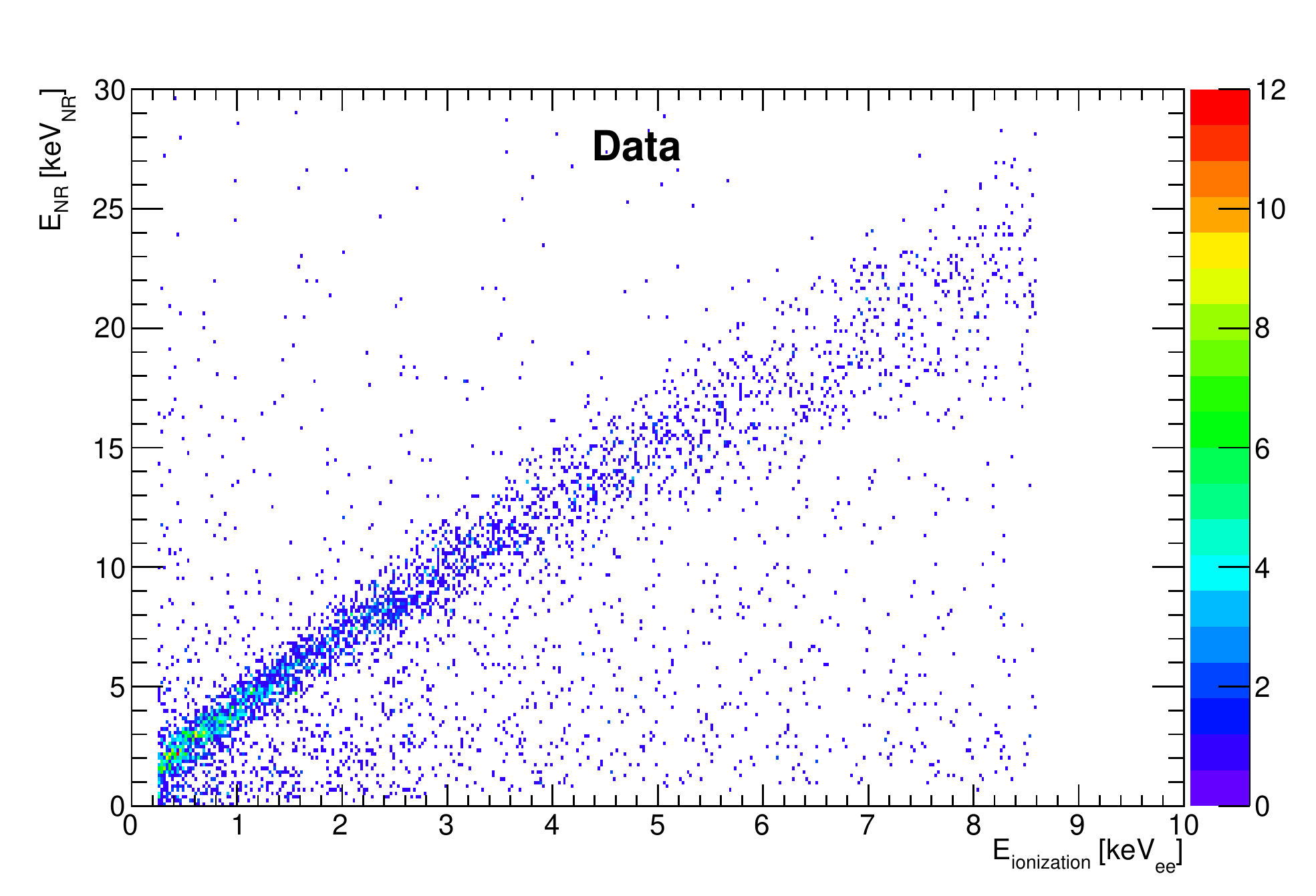}
		\includegraphics[width=0.7\textwidth]{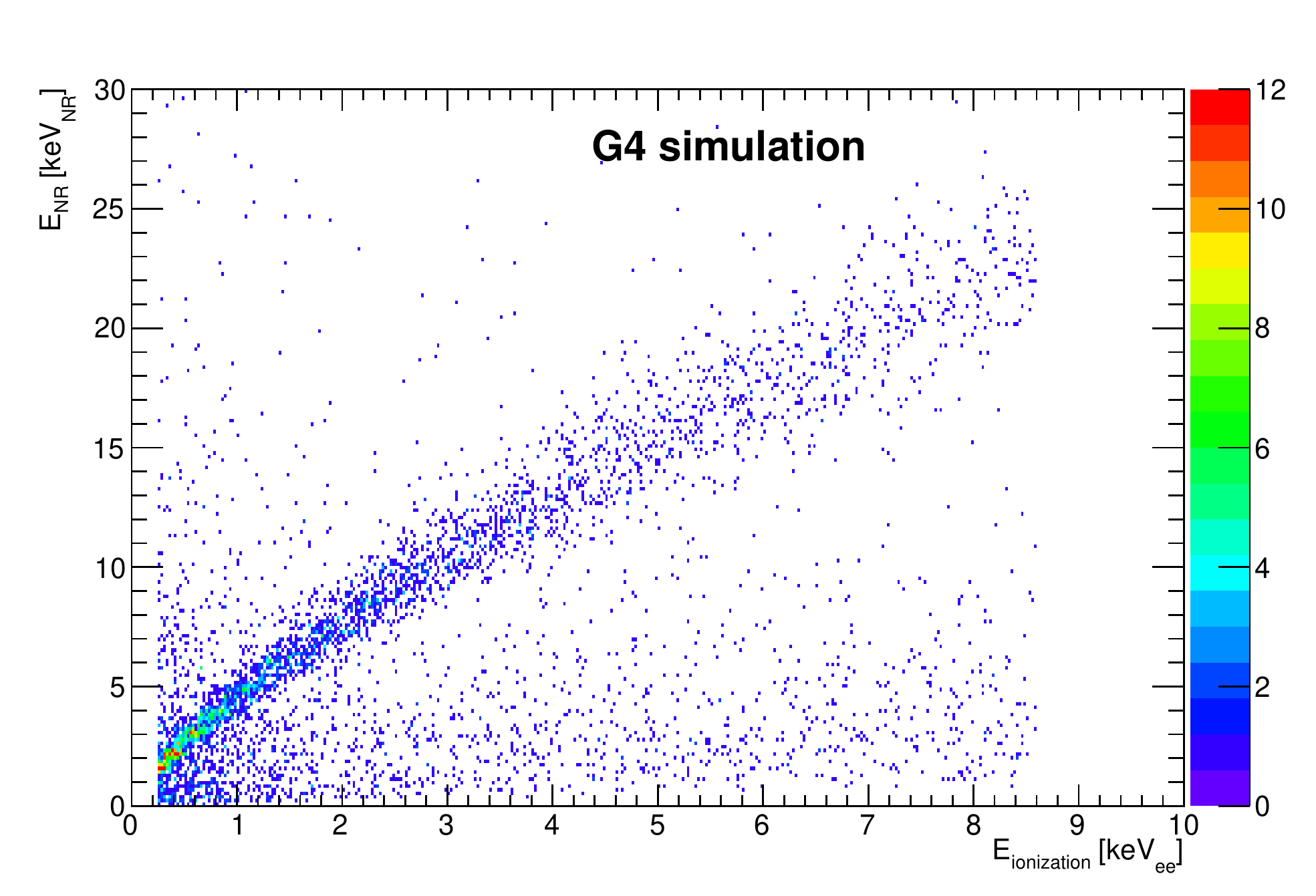}
		\caption{Energy of the nuclear recoil vs. energy in ionization including contributions of all the scintillator bars for the final data set (top panel) and for a simulated one with Geant4 (bottom panel).}	\label{fig:ErecVsEion}
	\end{center}
\end{figure}

About $1.5\times10^8$ events were recorded, most of which were noise in the SiDet. By requiring a coincidence such that both PMTs on a scintillator bar registering signals within 20~ns, the number of events was reduced to $1.8\times10^5$. Finally, after selecting events within the neutron arrival-time and eliminating events in which the energy was saturated beyond the dynamic range, $5.1\times10^3$ events survived. The 5,100 final events were placed in a single distribution, with contributions from all 21~bars. These events are shown in the top panel of figure~\ref{fig:ErecVsEion} as a color map. The accumulation of events in the diagonal of the plot is the signal while the sparse distribution is background. For comparison, the bottom panel of figure~\ref{fig:ErecVsEion} shows the result of the simulation. The spread in the signal band of the real data set has main contributions from the spread in recoil energy due to the finite size of the detectors which introduce an uncertainty in the scattering angle, the ToF resolution and fluctuations in the ionization yield from the nuclear recoil.  

\begin{figure}[t!]
	\begin{center}
		\includegraphics[width=0.7\textwidth]{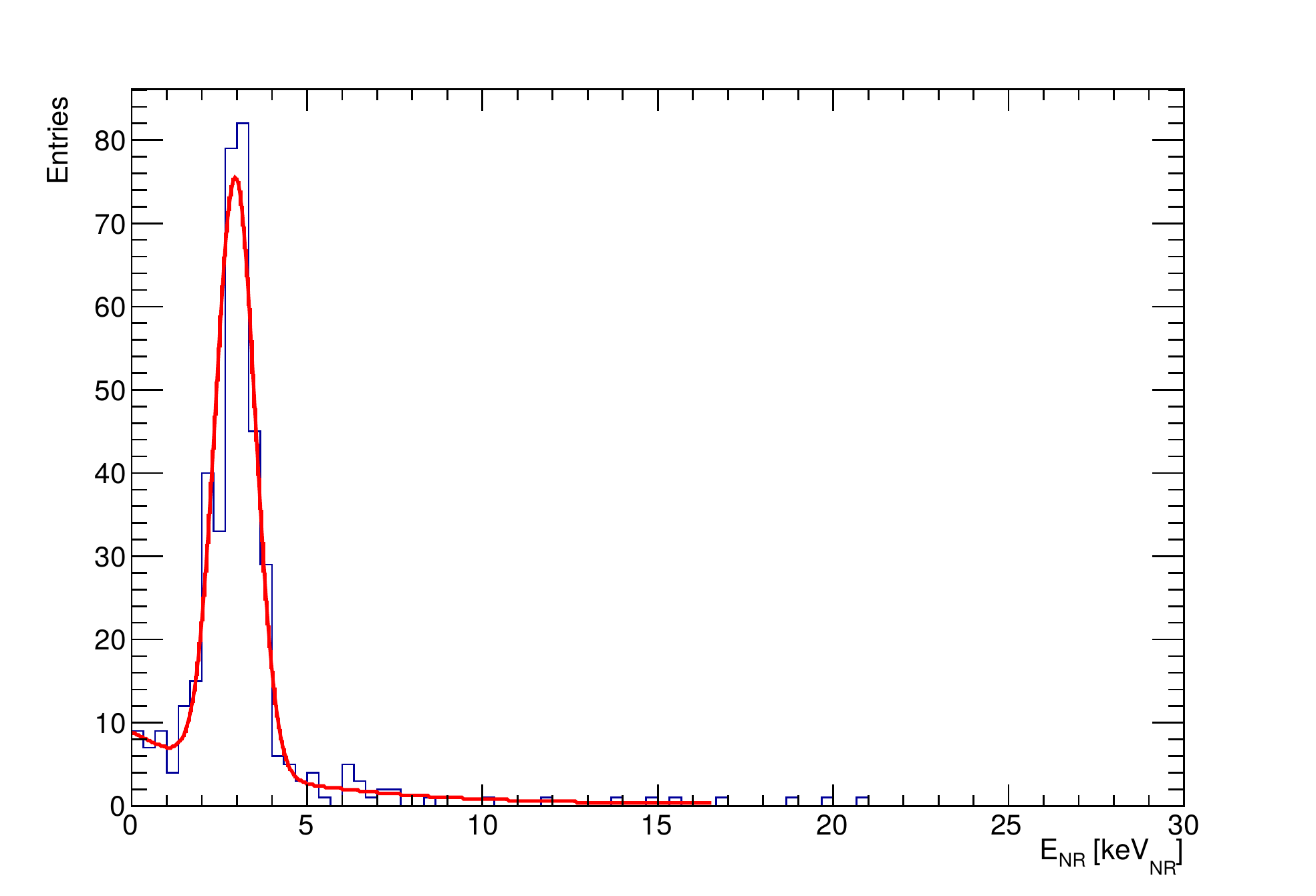}	\caption{Profile histogram of the top plot of figure~\ref{fig:ErecVsEion}. Distribution of $E_{NR}$, for $E_i$ in $[0.52, 0.76)$~keV$_\text{ee}$.}	\label{fig:ErecProfile}
	\end{center}
\end{figure}

The result was extracted using a binned-likelihood method. The horizontal scale was binned with variable bin size to account for the accumulation of events at low energies in ionization, such that the bin sizes are smaller at low energies. For each bin, a distribution of $E_{NR}$ was plotted, and fitted with a signal plus background function, where we used a Gaussian for the former and a decaying exponential for the latter, motivated by the simulation results. Figure~\ref{fig:ErecProfile} shows the profile histogram for $E_i$ of the ionization energy in the range from 0.52 to 0.76~keV$_\text{ee}$. The uncertainty returned by the fit for the mean of the Gaussian was taken as the statistical uncertainty in $E_{NR}$.

Systematic uncertainties from several sources were evaluated. The dominant uncertainty below $\approx 7$~keV$_{\mathrm{NR}}$ (2~keV$_{\text{ee}}$) is the uncertainty in the calibration of $E_i$ measured by the SiDet, described in section~\ref{sec:SiDet}. Above this energy, the uncertainty in the reconstruction of $E_{NR}$ dominates. It is affected by uncertainties in the geometry (the determination of $\theta$, $l$ and $r$) and the ToF measurement, where the largest contribution comes from the determination of $\theta$. 

To quantitatively evaluate these contributions simulations were used. Several effects due to possible geometric uncertainties were evaluated by modifying the absolute position and Euler angles of the components of the experiment. The effects studied included an offset in each coordinate in the position of the LiF target, the SiDet, a rigid offset of the whole setup, a rigid shift in the angles of the neutron detector array, the contribution of the relative orientation of the faces of the bars, and the contribution of a random change of all these effects on each individual bar. For evaluation, the components were displaced by 5 mm in the simulation, overestimating the accuracy of the surveyed geometry. 

The bias introduced by the extraction with the binned-likelihood method, i.e. the effect of events of low $E_{NR}$ propagating to higher energy bins because of the finite resolution in $E_i$ and the decreasing trend in event density with increasing $E_i$, was evaluated. It was found to be the dominant contribution in the determination of $E_{NR}$ of the two data bins of lowest energy, although the total systematic uncertainty on these two data points is dominated by the uncertainty in the determination of the SiDet energy scale. 

The effect of the neutron spectrum was also studied. Because the experimental technique presented in this work did not rely on the neutron beam energy spectrum, it was determined that even a major change in it would not introduce a systematic uncertainty. The effect of an increase and a decrease of the flux by a factor of two while keeping the spectrum constant, and of the use of a flat spectrum in [50,600]~keV was studied. The results obtained from these simulated data sets were consistent within the statistical uncertainties. 
 
\begin{figure}[t!]
	\begin{center}
		\includegraphics[width=0.7\textwidth]{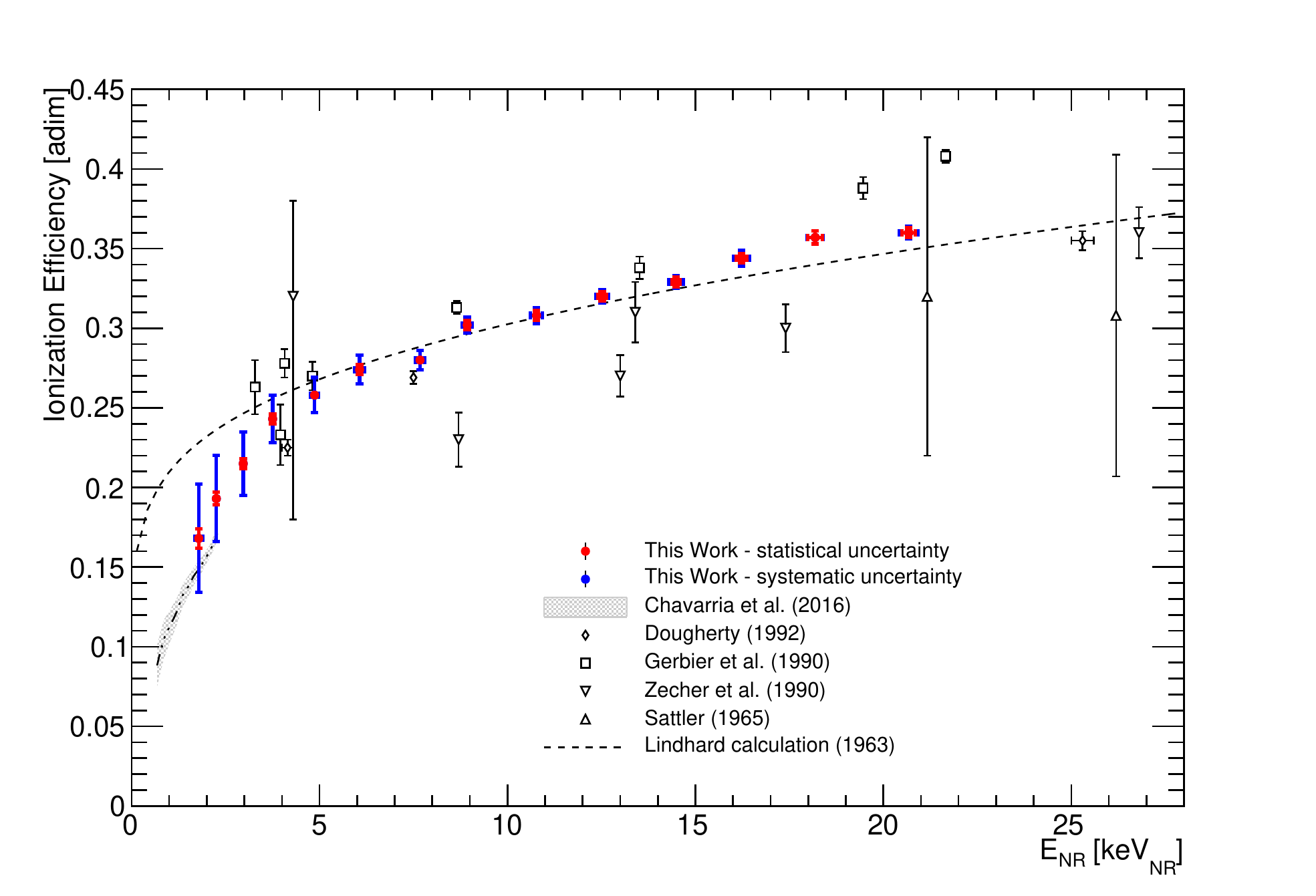}	\caption{Ionization efficiency (ratio between the energy released via ionization and the nuclear recoil energy) as a function of the nuclear recoil energy. The solid points are the result of this work, shown with the statistical (red) and systematic (blue) uncertainty bars. Also shown are data points from previous experiments: upward-pointing empty triangles from Sattler \cite{cite:Sattler}, downward-pointing empty triangles from Zecher \textit{et al.} \cite{cite:Zecher}, empty squares from Gerbier \textit{et al.} \cite{cite:Gerbier}, empty diamonds from Dougherty \cite{cite:Dougherty}, and grey area from Chavarria \textit{et al.} \cite{cite:UCresults}. The dashed curve is Lindhard prediction for silicon \cite{cite:Lindhard}. }	\label{fig:resultPlot}
	\end{center}
\end{figure}

Finally, the ionization efficiency, $\varepsilon = E_i/E_{NR}$, was calculated bin by bin. Table~\ref{tab:results} summarizes the results. Figure~\ref{fig:resultPlot} shows the result of this work compared to previous measurements and the Lindhard calculation. The overall trend of the presented measurement is well described by Lindhard theory above $\approx 4$~keV$_\mathrm{NR}$ of recoil energy. Below this energy the ionization efficiency measured drops faster than the model, confirming the deviation from the prediction observed in Ref. \cite{cite:UCresults}. This has a direct impact on dark matter and CENNS searches and their expected sensitivities at low energies, as such experiments typically use the Lindhard calculation down to detector-threshold energies which are often below 4~keV$_\text{NR}$.

\begin{table}[t!]
  \begin{center}
  \resizebox{\textwidth}{!}{
	\begin{tabular}{cccccccc}
	\hline \hline
	Energy in                &  Syst. unc.              & Nuclear Recoil           &  Stat. unc.              &  Syst. unc.              &  Ionization & Stat. unc.       & Syst. unc.       \\
        ionization, $E_i$        &  in $E_i$                &  Energy, $E_{NR}$        &   in $E_{NR}$            &    in $E_{NR}$           &  Efficiency & in $\varepsilon$ & in $\varepsilon$ \\
	$[\text{keV}_\text{ee}]$ & $[\text{keV}_\text{ee}]$ & $[\text{keV}_\text{NR}]$ & $[\text{keV}_\text{NR}]$ & $[\text{keV}_\text{NR}]$ &  [1]        &  [1]             &  [1]             \\ \hline
0.30  & $\pm$ 0.06 & 1.79  & $\pm$ 0.07 & $\pm$ 0.11 & 0.168 & $\pm$ 0.006 & $\pm$ 0.034   \\           
0.44  & $\pm$ 0.06 & 2.26  & $\pm$ 0.04 & $\pm$ 0.10 & 0.193 & $\pm$ 0.004 & $\pm$ 0.027   \\          
0.64  & $\pm$ 0.06 & 2.97  & $\pm$ 0.04 & $\pm$ 0.09 & 0.215 & $\pm$ 0.003 & $\pm$ 0.020   \\          
0.91  & $\pm$ 0.05 & 3.75  & $\pm$ 0.04 & $\pm$ 0.10 & 0.243 & $\pm$ 0.003 & $\pm$ 0.015   \\          
1.25  & $\pm$ 0.05 & 4.87  & $\pm$ 0.04 & $\pm$ 0.11 & 0.258 & $\pm$ 0.002 & $\pm$ 0.011   \\          
1.67  & $\pm$ 0.04 & 6.07  & $\pm$ 0.05 & $\pm$ 0.13 & 0.274 & $\pm$ 0.003 & $\pm$ 0.009   \\          
2.15  & $\pm$ 0.03 & 7.67  & $\pm$ 0.06 & $\pm$ 0.12 & 0.280 & $\pm$ 0.002 & $\pm$ 0.006   \\          
2.70  & $\pm$ 0.02 & 8.93  & $\pm$ 0.08 & $\pm$ 0.14 & 0.302 & $\pm$ 0.003 & $\pm$ 0.005   \\          
3.32  & $\pm$ 0.02 & 10.77 & $\pm$ 0.09 & $\pm$ 0.16 & 0.308 & $\pm$ 0.003 & $\pm$ 0.005   \\           
4.00  & $\pm$ 0.01 & 12.52 & $\pm$ 0.12 & $\pm$ 0.17 & 0.320 & $\pm$ 0.003 & $\pm$ 0.004   \\           
4.76  & $\pm$ 0.01 & 14.48 & $\pm$ 0.11 & $\pm$ 0.19 & 0.329 & $\pm$ 0.003 & $\pm$ 0.004   \\           
5.59  & $\pm$ 0.02 & 16.23 & $\pm$ 0.14 & $\pm$ 0.21 & 0.344 & $\pm$ 0.003 & $\pm$ 0.005   \\           
6.49  & $\pm$ 0.02 & 18.19 & $\pm$ 0.20 & $\pm$ 0.21 & 0.357 & $\pm$ 0.004 & $\pm$ 0.004   \\           
7.45  & $\pm$ 0.02 & 20.67 & $\pm$ 0.18 & $\pm$ 0.25 & 0.360 & $\pm$ 0.003 & $\pm$ 0.004   \\           
	\hline
	\end{tabular}
	  }
	\caption{Summary of the results. Measured Energy in ionization, $E_i$, as a function of the reconstructed Nuclear Recoil Energy, $E_{NR}$, with the corresponding uncertanties. }
	\label{tab:results}
  \end{center}
\end{table}

%%%%%%%%%%%%%%%%%%%
\section{Conclusions}

We presented a novel experimental method to measure the ionization energy produced by nuclear recoils from neutrons. Despite the fact that a neutron beam of a broad energy spectrum was used, the energy of each individual scattered neutron was measured using the neutron time of flight and scattering angle. This allowed for the reconstruction of the kinetic nuclear recoil energy of each interaction. The method was used to measured the ionization efficiency due to nuclear recoils in silicon between 1.8 and 20~keV$_\text{NR}$. The overall trend of the presented data is well described by Lindhard model above a recoil energy of 4~keV$_\text{NR}$. As energies decrease below this value, the data are found to have an increasingly lower ionization efficiency than is predicted by the model. The presented results are consistent with those of an independent experiment using a photo-neutron source \cite{cite:UCresults}, shown as the grey band in figure~\ref{fig:resultPlot}, confirming a significant deviation at low energies. This result impacts exclusion limits and future expected sensitivities of dark matter and CENNS searches that have used the Lindhard calculation for ionization efficiencies at low energies. This result can be used in conjunction with other measurements as a data-based prediction at such low energies.

%%%%%%%%%%%%%%%%%%%	
\section{Acknowledgements}

We would like to thank John Greene from Argonne National Laboratory for providing the LiF target used in the experiment. We thank to Fermilab staff, particularly to Andrew Lathrop and Rolando Flores for assistance in the neutron detector construction and deployment at UND, the Meson Assembly Building machine shop technicians, and the Alignment and Metrology Department members O'Sheg Oshinowo, Charles Wilson and Michel O'Boyle. This work was supported by the National Science Foundation under Grant No. PHY-1419765. We thank the CONACYT from Paraguay for the support by the project 14-INV-092. Support from grant 203501 of CONACYT, M\'{e}xico, is acknowledged. This work was supported by the Swiss National Science Foundation grant 153654.

%%%%%%%%%%%%%%%%%%%

\end{document}